\documentclass[a4paper,8pt]{article}
\usepackage[cp1251]{inputenc}
\usepackage[russian,english]{babel}
\usepackage{array,longtable}
\usepackage{amssymb, latexsym, amsthm, amsmath, amsxtra}
\usepackage{graphicx}

\graphicspath{{}}
\DeclareGraphicsExtensions{.pdf,.png,.jpg,}

\makeatletter

\begin{document}
\large

\begin{center}
\title{}{\bf  On the analogy between consequences of the breather model of particles interaction and anomalous flow of Helium-II. }
\vskip 1cm

\author{}
 R.K. Salimov \textsuperscript{1}, E.G. Ekomasov \textsuperscript{1}
{}

% author(-s)
\vskip 1cm

{ \textsuperscript{1} Bashkir State University, 450076, Ufa, Russia }

%The name of establishment in which research is executed.
%If authors take an identical place of work, the name of the organization is written once.

\vskip 0.5cm
e-mail: salimovrk@bashedu.ru

\end{center}

\vskip 1cm

{\bf Abstract}
 \par
 The article deals with the sin-Gordon model and considers the interaction of breathers localized on moving impurities. Numerical simulation of collective behavior of such objects revealed the analogy between the model’s properties and the properties of Helium-II anomalous flow. 

 \par
 \vskip 0.5cm

{\bf Keywords}:  nonlinear differential equations,  breather, wave-particle duality, superfluidity, Helium-II anomalous flow.

\par
\vskip 1cm

{\bf 1.Introduction}
\vskip 0.5cm

Soliton solutions of nonlinear equations such as the sin-Gordon equation are often used as models of stable extended particles [1]. In particular the Skyrme model for barions can be noted [2,3]. In such a case stationary solutions for such equations are used as particles.  Nonlinear equations, including the sin-Gordon equation, have breather solutions as well. Breather solutions for Lorentz-invariant nonlinear equations are engaging since in their moving there is space modulation analogous to de Broglie wave. Indeed, it is possible to write the following expression for a moving breather of the sin-Gordon equation :
   \begin{align}
   u=4\arctg(\frac{a \sin(b(t-vx)/\sqrt{(1-v^2)})}{b \cosh(a(x-vt)/\sqrt{(1-v^2)})}) ,\label{eq:1}
   \end{align}
Where u is the solution in a breather form, and to then obtain a space modulation:
 \begin{align}
  sin(b(t-vx)/\sqrt{(1-v^2)}) ,\label{eq:2}
   \end{align}

analogous to the de Broglie wave modulation

\begin{align}
  \lambda=\frac{h}{mv},\label{eq:3}
   \end{align}

In such a case the particle mass in the de Broglie formula can be associated with the frequency of breather vibrations. The disadvantage of such models is, firstly, the absence of exact breather solutions for the three-dimensional case [4]; secondly, the discrepancy between  point size of particles and extended size of breathers. To overcome the disadvantages of the particles breather models a “model of moving impurities” can be suggested. Impurities creating potential holes for some scalar field with breathers in them will act as analogues of particles. A breather in such a case can be considered as some pilot wave [5] for the impurity.  An example of such a model of particles can be the so-called “walkers” [6] for which the role of a particle is played by a drop and the role of a pilot wave – by the wave around it. Notwithstanding the uncommonness of the model we further on show that the collective behavior of such objects can have interesting consequences.

                            {\bf 2.Main equations and results.}

\vskip 0.5cm

Let us consider the sin-Gordon equation as an example for the Hamiltonian model [7] modifying the equation for the two-dimension case in the presence of a moving impurity:

 \begin{align}
 \frac{ r_{0t}^2}{2}+\int(\frac{u_x^2}{2}+\frac{u_y^2}{2}+\frac{u_t^2}{2}+2\sin^2(\frac{u}{2})f(r_0,x,y))dxdy,\label{eq:4}
   \end{align}

where  $r_0$ are the coordinates of a moving nonhomogeneity with its own unit mass, $f(r_0,x,y)$  is the potential hole created by the nonhomogeneity. Expression (4) is a common Hamiltonian for the sin-Gordon equation field with impurity. The only difference is the impurity mobility and taking into account its kinetic energy. The two-dimensional case was chosen as the least time-consuming for calculations and it allows us to model the effects under consideration. (4) makes it possible to obtain the following motion equations for the field:

\begin{align}
 u_{xx}+u_{yy}-u_{tt}-sin(u)f(r_0,x,y)=0,\label{eq:5}
   \end{align}

And for the coordinates of the impurity center:

\begin{align}
 x_{0tt}=-2\int\sin^2(u/2)\frac{\partial f(r_0,x,y)}{\partial x_0} dxdy,\label{eq:6}
   \end{align}

\begin{align}
 y_{0tt}=-2\int\sin^2(u/2)\frac{\partial f(r_0,x,y)}{\partial y_0} dxdy,\label{eq:7}
   \end{align}

Besides a Hamiltonian (4) a similar vector model with two degrees of freedom can be considered
\begin{align}
H= \frac{ r_{0t}^2}{2}+\int(\frac{u_x^2}{2}+\frac{u_y^2}{2}+\frac{u_t^2}{2}+2\sin^2(\frac{u}{2})+\frac{v_x^2}{2}+\frac{v_y^2}{2}+\frac{v_t^2}{2}+2\sin^2(\frac{v}{2})+\alpha sin(u^2+v^2))dxdy,\label{eq:8}
   \end{align}
In the process of numerical simulation a difference scheme used to study the sin-Gordon equation [8] was involved. Picture 1 shows the movement of the breather and the impurity for the model described by expression (8). The function for the impurity has the form of a potential hole.When moving the breather displays the presence of space modulation. The degree of the black color gradation in the picture is proportional to the magnitude of the field size  . In the picture the impurity center creating a potential hole is marked by a square. Let us point out that such a common movement of the impurity and breather is rather interesting from the point of view of modelling wave effects such as interference and tunneling.

\begin{figure}[!ht]
\center
\includegraphics[width=6.5cm, height=3.5cm]{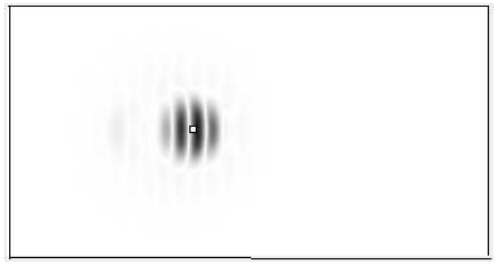}
\caption{Movement of the breather and the impurity connected to it.}
\label{schema}
\end{figure}

Let us consider collective behavior of such objects. Numerical simulation of generalization of equations (5)-(7) for two impurities shows that interaction of such breathers and impurities will depend on mutual difference of vibration phases – if the breather vibrates within the phase then the impurities attract; the impurities repel in the reversed phase. For example in the case of
\begin{align}
f(r_{0,i},x,y)=-\cos(((x-x_{0,i})^2+(y-y_{0,i})^2)\pi/9)-1;r_{0,i}-r<3;r_{0,1}-r_{0,2}=8,\label{eq:6}
   \end{align}

The phase difference in reversed phase is $\pi$ . Since it is time-consuming to numerically consider collective behavior of many breathers, we simplify the problem and observe the particles whose interaction depends on some phase. In such a case we assume that the shape and frequency of breathers do not alter during the interaction, and we only consider the motion equations for the coordinates of impurities. Further on we assume that moving impurities will interact the same way material particles do. The difference of such breather interaction from the interaction of material particles is in the dependence on the vibration phase. As an example let us numerically consider, using the method of molecular dynamics, the set of particles whose interaction depends not only on the distance between them but also on the difference of phases. For that we in a random manner ascribe to each particle some initial phase $\phi$  of vibrations and will assume that it does not alter over time. The vibration phases will take on values from $0$  to $2\pi$  , the distribution is uniform.
First  we consider 144 particles pair potential of which takes the form for $t=0$ .

\begin{align}
 V(r_{ij})=\frac{28}{1+r_{ij}^4}-2\exp(-0.28(r_{ij}-2)^2)         ,\label{eq:9}
   \end{align}
Such a potential does not limit generality of consideration and is analogous to Lennard-Jones potential [9] since it has a minimum and   decays away rather fast (fig. 2).

\begin{figure}[!ht]
\center
\includegraphics[width=7cm, height=4cm]{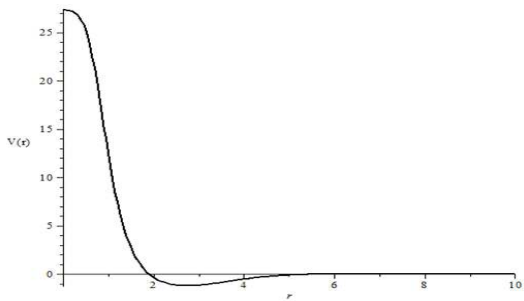}
\caption{. Potential type (10).}
\label{schema}
\end{figure}

\begin{figure}[!ht]
\center
\includegraphics[width=13cm, height=4.5cm]{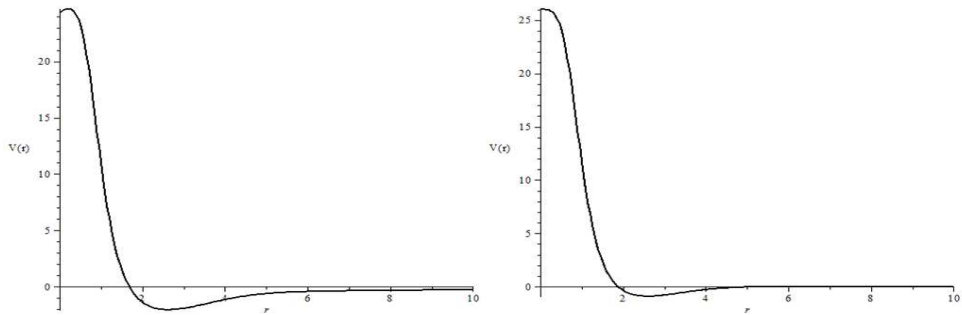}
\caption{. Potential type (11).}
\label{schema}
\end{figure}

Here  - $r_{ij}$ is the distance between $i$  and $j$  particles. The other potential  for taking account of the interaction depending on the phase difference has the form

\begin{align}
 V(r_{ij})=\frac{28}{1+r_{ij}^4}-2\exp(-0.28(r_{ij}-2)^2)+\frac{1.2\cos(\phi_i-\phi_j)}{1+r_{ij}}       ,\label{eq:10}
   \end{align}

Here $\phi_i$-  is phase of $i$  particle. Fig. 3 represents the form of the potential of interaction (11)  for  in-phase and reversed phase cases.

Motion equations of molecular dynamics  are deduced from the potential as follows:

\begin{align}
 x_{itt}=-\sum_{i \neq j}\frac{\partial V}{\partial r_{ij}}\frac{\partial r_{ij}}{\partial x_j},\label{eq:11}
   \end{align}

\begin{align}
 y_{itt}=-\sum_{i \neq j}\frac{\partial V}{\partial r_{ij}}\frac{\partial r_{ij}}{\partial y_j},\label{eq:12}
   \end{align}

Let us assume the initial location of the particles to be their matrix-like arrangement as showed in fig. 4.

\begin{figure}[!ht]
\center
\includegraphics[width=14cm, height=12cm]{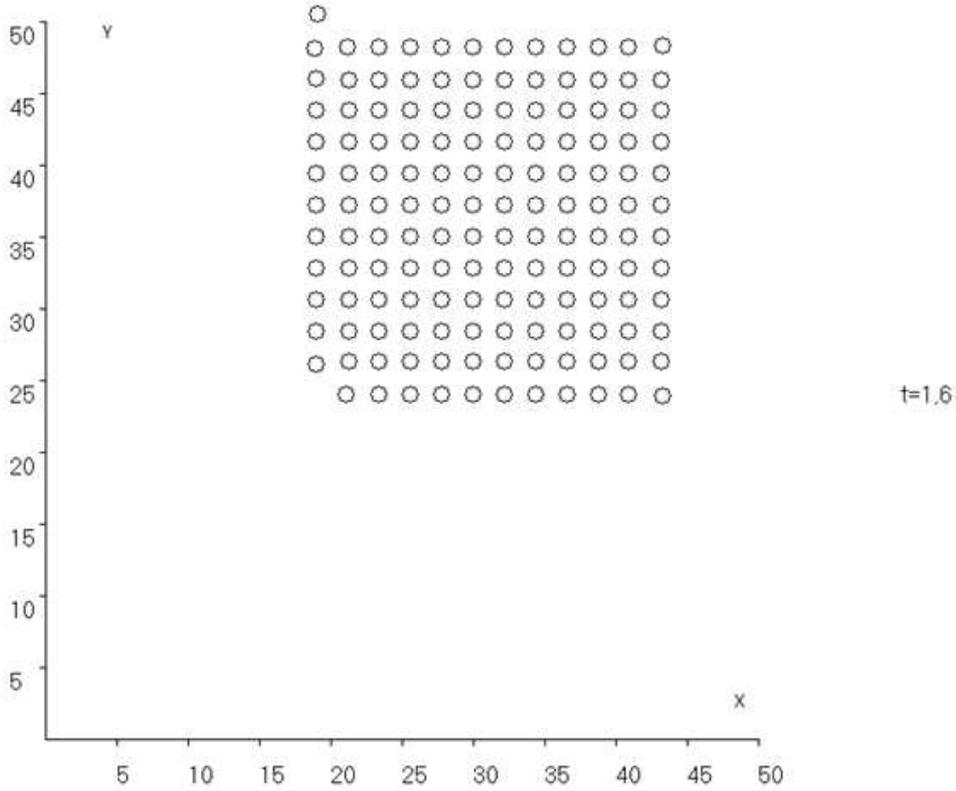}
\caption{Initial location of the particles}
\label{schema}
\end{figure}

Then during numerical simulation velocities of the particles are multiplied by 0,99995 for each time step ($dt=0.01$ ). In such a case the particles velocity does not drop to zero during the period of monitoring. Runge-Kutta method of the fourth order was used for numerical simulation [10]. Over time for potential (10) we obtain the stable condition showed in fig. 5.
\begin{figure}[!ht]
\center
\includegraphics[width=13cm, height=11cm]{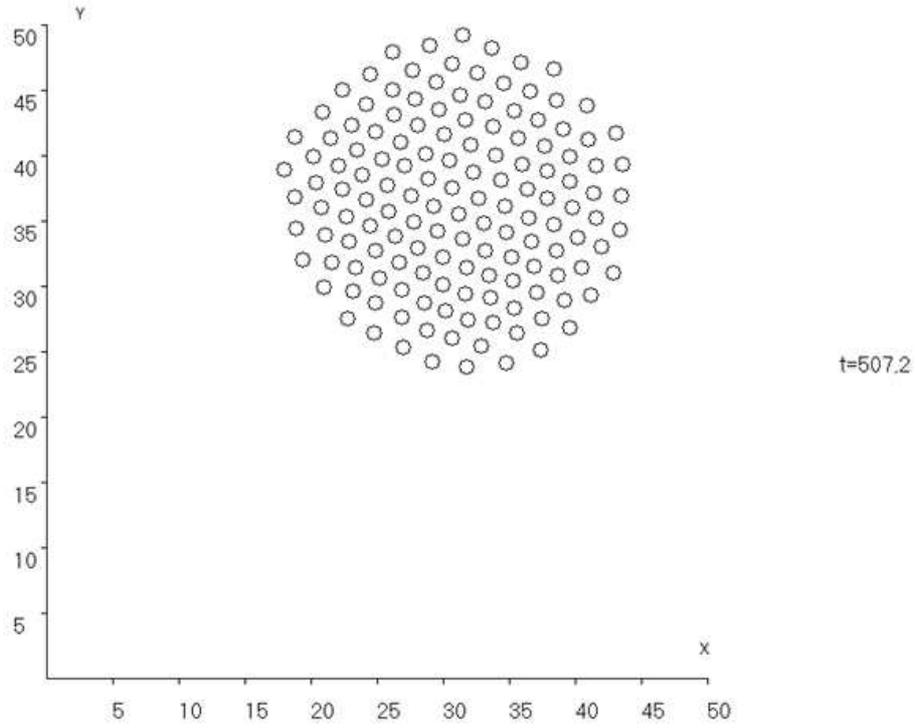}
\caption{ Stable condition for potential (10). }
\label{schema}
\end{figure}

For potential (11) in which the interaction depending on the phase slowly linearly turns on from  0 to value 1,2  ($t<400$  ), simultaneously with the reduction of particles velocity (“or cooling”)  over time we obtain the following conditions showed in fig. 6. For other initial random conditions the situation is the same.

\begin{figure}[!ht]
\center
\includegraphics[width=14cm, height=11cm]{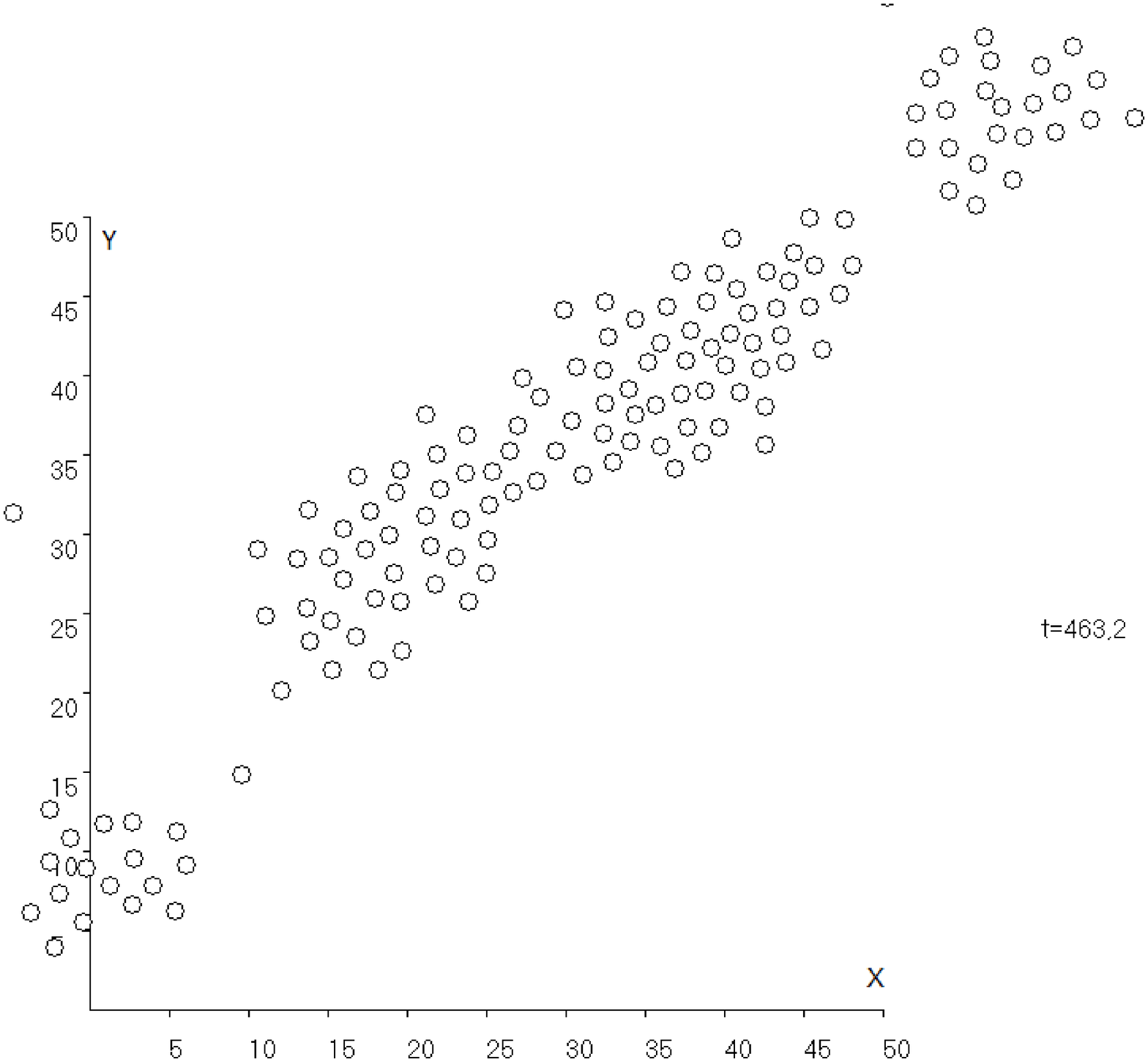}
\caption{ Dispersing of particles under potential impact (11), t=463. }
\label{schema}
\end{figure}

As it can be seen in the pictures when the interaction depending on the phase turns on, the particles tend to arrange in an extended form thus expanding the perimeter of the border of particles conglomerate, which we will further on refer to as “liquid”. This happens due to the fact that particles in the reversed phase repel. At such interaction the particles are distributed in such a way that particles with similar phases happen to locate close to each other. Linear sizes along the extended direction outdo the linear sizes of roughly round state described by potential (10). It is interesting to note that such a property of the model is analogous to the properties of anomalous flow of Helium II which in superfluid state, as it is know from experiments, flows out of reservoirs  [11,12].
One more property of the “breather model” – the increase of energy of connection between closely located particles – can be associated with termination of boiling [11,12]  at transition into Helium II. Indeed, the force between the particles interacting in the  in-phase in the state described by potential (11) will be greater than for potential (10) thus the energy needed for removal of the particle from the rest of the “liquid” will be greater as well. Hence the “boiling temperature” of such “liquid” will also be higher and the property is analogous to the termination of boiling of Helium II. The change of the value and the character of the  potential of interaction of particles with the same phase in a breather model are similar to strength enhancement of nanodroplets in superfluid helium [13].

   {\bf 3.Conclusion.}

The model presented above is an effort to apply breather interactions from the point of view of quasimicroscopy. It has been showed that being uncommon the model of interaction of the particles, whose pair interaction depends on some vibration phase, does demonstrate rather interesting results. For instance, it qualitatively describes some properties analogout to those of Helium II behavior. The authors find it challenging since, although the nature of superfluidity has been studied thoroughly, it has only been explained by macroscopic quantum theories [14-16].

%********************Bibliography.*******************************************************

\end{document}